\documentclass[sigconf]{acmart}

\AtBeginDocument{%
  \providecommand\BibTeX{{%
    \normalfont B\kern-0.5em{\scshape i\kern-0.25em b}\kern-0.8em\TeX}}}

\copyrightyear{2024} 
\acmYear{2024} 
\setcopyright{rightsretained} 
\acmConference[WWW '24 Companion]{Companion Proceedings of the ACM Web Conference 2024}{May 13--17, 2024}{Singapore, Singapore}
\acmBooktitle{Companion Proceedings of the ACM Web Conference 2024 (WWW '24 Companion), May 13--17, 2024, Singapore, Singapore}\acmDOI{10.1145/3589335.3651251}
\acmISBN{979-8-4007-0172-6/24/05}

\acmSubmissionID{de9269}

\usepackage{caption}
\usepackage{subcaption}

\makeatletter
\gdef\@copyrightpermission{
  \begin{minipage}{0.3\columnwidth}
   \href{https://creativecommons.org/licenses/by/4.0/}{\includegraphics[width=0.90\textwidth]{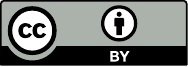}}
  \end{minipage}\hfill
  \begin{minipage}{0.7\columnwidth}
   \href{https://creativecommons.org/licenses/by/4.0/}{This work is licensed under a Creative Commons Attribution International 4.0 License.}
  \end{minipage}
  \vspace{5pt}
}
\makeatother

\begin{document}

\title{SocialGenPod: Privacy-Friendly Generative AI Social Web Applications with Decentralised Personal Data Stores}






\author{Vidminas Vizgirda}
\email{s1750767@ed.ac.uk}
\orcid{0000-0002-5085-9513}
\affiliation{%
  \institution{University of Edinburgh}
  \streetaddress{}
  \city{Edinburgh}
  \state{}
  \country{UK}
  \postcode{}
}

\author{Rui Zhao}
\email{rui.zhao@cs.ox.ac.uk}
\orcid{0000-0003-2993-2023}
\affiliation{%
  \institution{University of Oxford}
  \streetaddress{}
  \city{Oxford}
  \state{}
  \country{UK}
  \postcode{}
}

\author{Naman Goel}
\email{naman.goel@cs.ox.ac.uk}
\orcid{0000-0002-5106-5889}
\affiliation{%
  \institution{University of Oxford}
  \streetaddress{}
  \city{Oxford}
  \state{}
  \country{UK}
  \postcode{}
}

\renewcommand{\shortauthors}{Vidminas Vizgirda, Rui Zhao, and Naman Goel}

\begin{abstract}
   We present SocialGenPod, a decentralised and privacy-friendly way of deploying generative AI Web applications. Unlike centralised Web and data architectures that keep user data tied to application and service providers, we show how one can use Solid --- a decentralised Web specification --- to decouple user data from generative AI applications. We demonstrate SocialGenPod using a prototype that allows users to converse with different Large Language Models, optionally leveraging Retrieval Augmented Generation to generate answers grounded in private documents stored in any Solid Pod that the user is allowed to access, directly or indirectly. SocialGenPod makes use of Solid access control mechanisms to give users full control of determining who has access to data stored in their Pods. SocialGenPod keeps all user data (chat history, app configuration, personal documents, etc) securely in the user's personal Pod; separate from specific model or application providers. Besides better privacy controls, this approach also enables portability across different services and applications. Finally, we discuss challenges, posed by the large compute requirements of state-of-the-art models, that future research in this area should address. Our prototype is open-source and available at: \url{https://github.com/Vidminas/socialgenpod/}.
\end{abstract}





\maketitle

\section{Introduction}
Many Web applications using generative AI like large language models (LLMs)~\cite{zhao2023survey} and diffusion models~\cite{yang2023diffusion} have emerged recently. Applications like \href{https://chat.openai.com}{ChatGPT}, \href{https://openai.com/dall-e-2}{DALL-E}, \href{https://www.perplexity.ai/}{Perplexity}, \href{https://www.maket.ai/}{Maket.ai}, \href{https://www.synthesia.io/}{Synthesia}, \href{https://writesonic.com/}{WriteSonic}, \href{https://www.jasper.ai/}{Jasper} are plentiful in domains such as search, writing, music, and design. Such applications usually manage the data of their users in a centralised way. For example, in a chat application, data from different users (such as user queries and answers generated by an LLM) are managed by the chat application provider, usually using a centralised database. Users have little control over how this personal data is used by the application provider~\cite{la2023language}. Centralised storage of sensitive user data also raises other privacy concerns. For example, in March 2023, a technical glitch in ChatGPT allowed some users to see parts of other users' conversations~\cite{bbc-gptchatleak}.

\begin{figure}
  \centering
  \includegraphics[width=\linewidth]{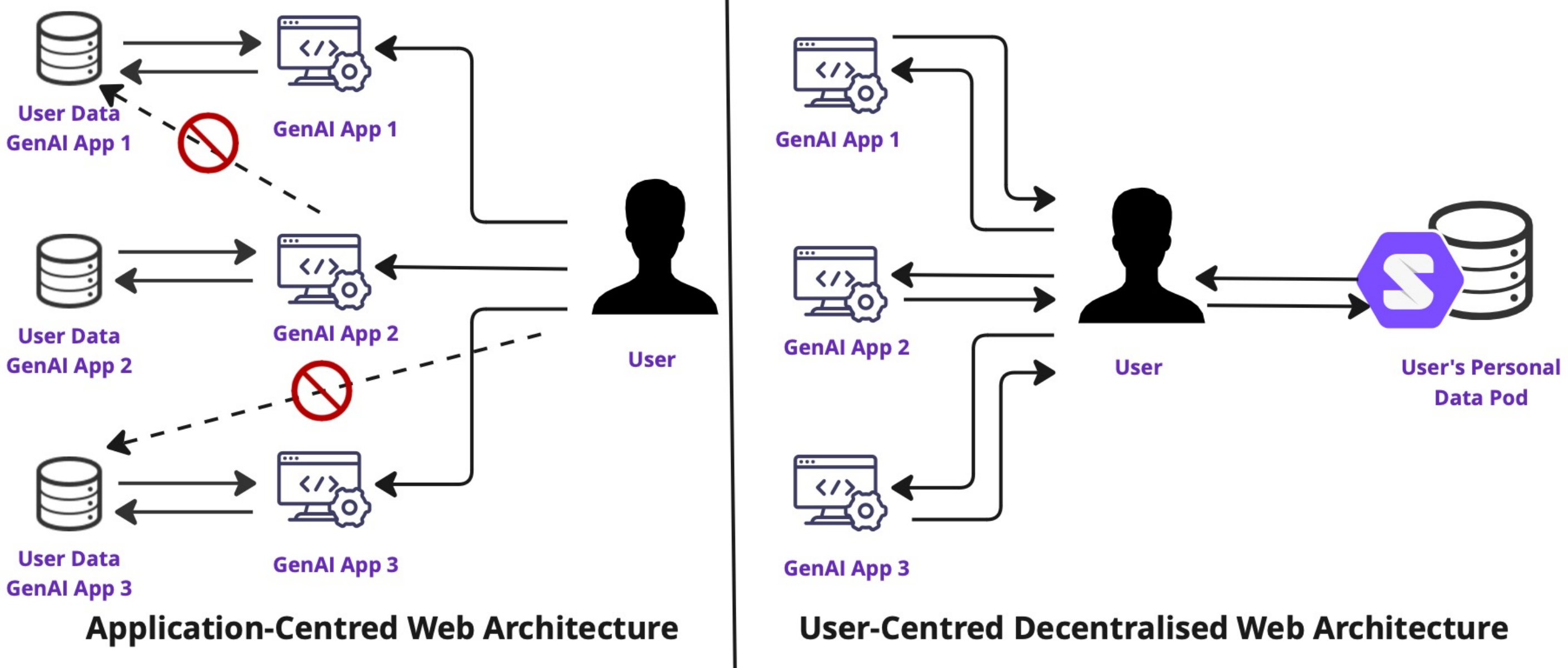}
  \caption{Centralised architecture (left) makes it difficult for users to have control over their data and use their data across applications. Decentralised architecture (right) decouples data from applications, giving users control over their data and making it easier for them to use their data across apps.}
  \Description[Side by side graphics depicting databases, apps, users, and arrows between them]{Side by side architecture illustrations. On the left, a user uses 3 different apps, each of which has its own data storage. The user cannot access these data stores directly, and the apps cannot access different apps' data. On the right, 3 different apps all communicate with a Solid Pod via the user.}
  \label{fig:fig1}
\end{figure}

Other than control and privacy concerns, another drawback of centralised and application-specific data management is that users cannot easily re-use their data from one application in another (also known as the vendor lock-in problem). For example, a user may not be able to use their data easily from an application for conducting background research in another application for writing or preparing graphics. Ideally, people should be able to use their data in the best applications available in the market, without unnecessary obstructions. This is illustrated in \autoref{fig:fig1}. 

To address these issues, we demonstrate a decentralised and privacy-friendly way of deploying generative AI Web applications. The key idea in this work is to use decentralised personal data stores for managing user data as illustrated on the right of \autoref{fig:fig1}. Specifically, we use Solid~\cite{mansour2016demonstration, sambra2016solid}, a decentralised Web specification based on standard, open, and interoperable data formats and protocols. It allows users to store their data in their Web-accessible Pods (Personal Online Data Stores) and configure granular access control for applications and other Pod/WebID owners. Solid also handles secure transmission of data for authorised requests. To support data portability, Solid apps and services read and write data to the users' Pods instead of siloed app backends. Users can transfer their data from one Pod service provider to another, or even host one themselves.
But there are also several technical challenges in using Solid with generative AI models. One challenge is the large computation power required for running these models, which makes an entirely decentralised and privacy guaranteeing implementation difficult. While user queries and generated responses can be stored in a user's Pod, large AI model inference cannot run locally (for example, in browser) with reasonable performance on typical end-user devices. Another challenge for RAG (Retrieval Augmented Generation) --- a technique which uses personal documents stored in user Pods for grounding the responses --- is that similarity search is required to pick out relevant documents. Existing Solid server implementations do not offer any support for similarity search, neither with vector embedding techniques nor other frequently used approaches, like Okapi BM25. 

In the paper, we discuss how we overcome some of these challenges for building our prototype and future research directions that will be of interest to the attendees of TheWebConf. Attendees will also be able to interact with the prototype. Interested community members can also contribute to further development of the open-source prototype.

\section{Related Work}\label{sec:related}

There are several related approaches for decentralised and privacy-friendly deployment of generative AI models. These include training highly capable small models (e.g., models with fewer parameters, distilled models, quantised models, etc) that can run with less compute and making the models openly available. Examples include the LLaMA models from Meta~\cite{touvron2023llama}, MoE models from Mistral AI~\cite{jiang2024mixtral}, and many others available on the open-source hub \href{https://huggingface.co/models}{HuggingFace}.


Several emerging applications rely on local generative AI models to limit data privacy concerns with retrieval augmented generation. For example, \href{https://www.rewind.ai/}{Rewind} gathers sensitive user data, such as screen recordings, emails, and meeting summaries, to enable searching past user activity. Their approach is to minimise data that leaves the user's computer by using local device resources for AI model inference \cite{rewind-faq}. Similar ideas are also used by projects such as \href{https://webllm.mlc.ai}{WebLLM},  \href{https://github.com/imartinez/privateGPT}{PrivateGPT} and \href{https://github.com/marella/chatdocs}{ChatDocs}. However, running models locally is often infeasible on end-user devices due to the high compute requirements of large AI model inference. Local models are also not sufficient in use cases that require data sharing between users.

We also note the ``AI gateway" approach, such as the one implemented by \href{https://github.com/wealthsimple/llm-gateway}{llm-gateway}. AI gateways act as a proxy between users and hosted models, intercepting user data before it is sent to external models, and scrubbing personal information or performing other security checks. This approach also avoids vendor lock-in by providing a unified API for interacting with different services. While there is an overlap in the motivation, the AI gateway approach focuses on single-user single-application functionalities, without features for collaboration between applications and users.

Tim Berners-Lee recently published a design note~\cite{privatedata}. There are some similarities in the ideas proposed in his design note and SocialGenPod. But, to the best of our knowledge, SocialGenPod is the first concrete implementation of RAG using Solid.

\section{Overview of \NoCaseChange{SocialGenPod}}

\begin{figure}[h]
  \centering
  \includegraphics[width=\linewidth]{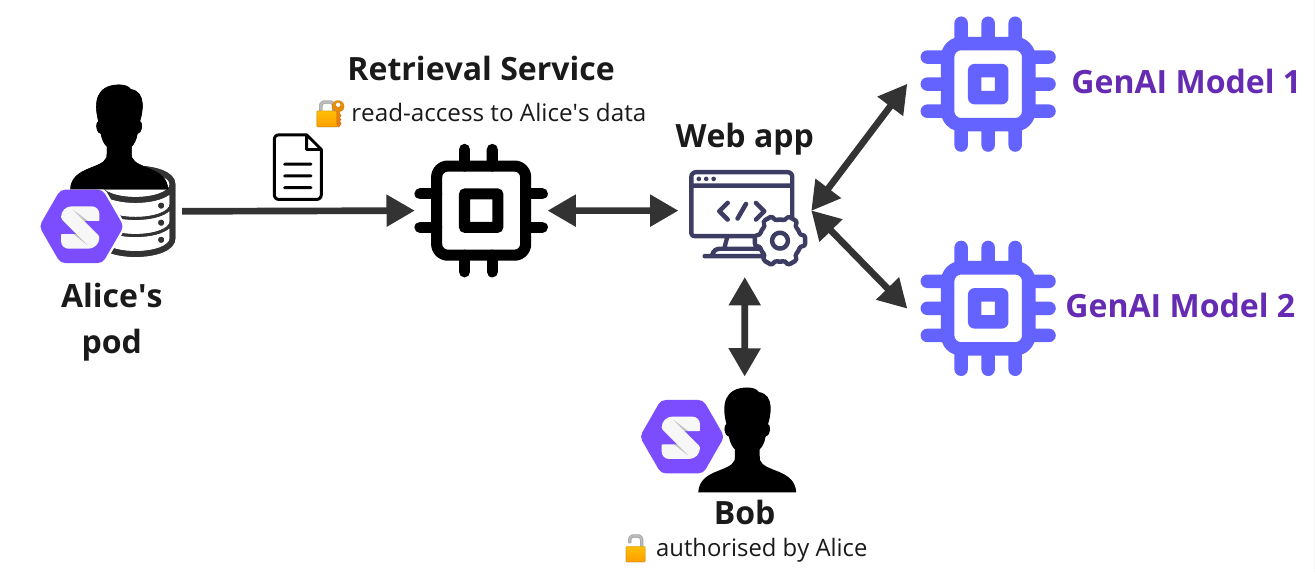}
  \caption{Example Web application data flow in SocialGenPod. Alice allows Bob to interact with or query her personal data through a Web app. The app uses a retrieval service to find permitted documents in Alice's Solid Pod that are most relevant to Bob's query. The app then uses one of the available generative AI models to generate a contextualised response for Bob.}
  \Description[Data flow graphic]{Data flow graphic that depicts Bob (with note ``authorised by Alice'') interacting with a Web app. The Web app interacts with a Retrieval Service (with note ``read-access to Alice's data) which obtains documents from Alice's pod. The Web app also interacts with GenAI Model 1 and GenAI Model 2.}
  \label{fig:fig-zoom}
\end{figure}

\begin{figure}[t]
  \centering
  \begin{subfigure}{\linewidth}
    \includegraphics[width=\linewidth]{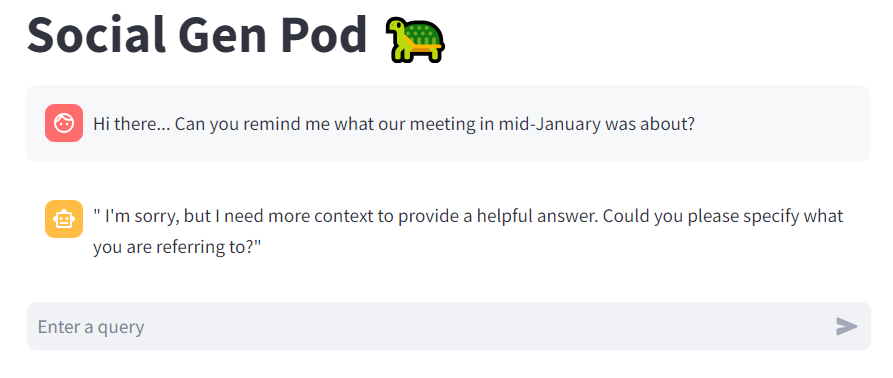}
    \caption{Without RAG}
  \end{subfigure}
  \begin{subfigure}{\linewidth}
    \includegraphics[width=\linewidth]{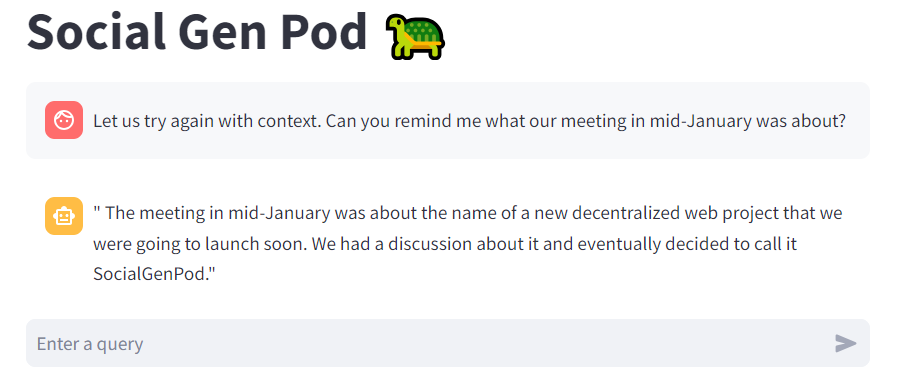}
    \caption{With RAG}
  \end{subfigure}
  \caption{RAG with private data in SocialGenPod. A chatbot without RAG (top) is unable to answer a user's query. The same chatbot with RAG (bottom) produces a better response.}
  \Description[Stacked screenshots of 2 chats]{Stacked screenshots of 2 SocialGenPod chats. In the top one labelled ``Without RAG'', the user wrote ``Hi there... Can you remind me what our meeting in mid-January was about?'' and the chatbot responded ``I'm sorry, but I need more context to provide a helpful answer. Could you please specify what you are referring to?''. In the bottom screenshot labelled ``With RAG'', the user wrote ``Let us try again with context. Can you remind me what our meeting in mid-January was about?'' and the chatbot responded ``The meeting in mid-January was about the name of a new decentralised web project that we were going to launch soon. We had a discussion about it and eventually decided to call it SocialGenPod.''}
  \label{fig:fig-ui}
\end{figure}

\textbf{Example Scenario:} Let us consider an example use-case scenario (\autoref{fig:fig-zoom}) that requires Retrieval Augmented Generation (RAG) and user data sharing. Suppose there are two friends or collaborators: Alice and Bob. Alice stores personal documents, such as her notes on a particular project, in her Solid Pod. She then configures a virtual ``personal AI assistant'' to have read access to this data. The virtual personal assistant is a Web app with Alice's configuration and is powered by an embeddings model for retrieval and a Large Language Model (LLM). The models may be running either on Alice's machine or on an external service. Bob can chat with this virtual assistant (using the Web app) and learn from Alice's project notes (potentially avoiding scheduling an unnecessary meeting), without ever getting a copy of the full documents. Alice could also configure access permissions so only a specific set of friends or users could interact with her personal AI assistant. \autoref{fig:fig-ui} shows SocialGenPod prototype using RAG with private data.

In the above example, the defining feature of Alice's personal AI assistant is Alice's configuration of the chatbot Web app, i.e., which service providers it uses and where it sources Alice's documents from. The AI models themselves are not necessarily personalised --- it is access to Alice's documents that makes it a personal assistant. The Web app itself could run anywhere (for example, Alice could host it or share the source code for others to run or point to a trusted public app provider, run by, e.g., a government or union).

The above use-case example can be generalised to other domains where private data sharing is relevant, for example, commercial consulting, financial advice, medical advice, career advice, meeting scheduling, or personalised feedback. It does not have to be restricted to just language generation either --- the same scenario could apply to multi-modal models too. The data stored in Solid Pods can also include also other kinds of data such as chat history between two users, chat history between a user and an LLM, etc. Solid Pods store all these data in a privacy-friendly and user-centric manner, and subject to user specified preference, can be used by various innovative and useful generative AI applications (personal AI assistants being one example).


\section{Design and Usage Details}

SocialGenPod is built using Solid, leveraging its mechanisms for access control, user authentication, and data discovery. As shown in \autoref{fig:fig-zoom}, SocialGenPod contains several modular components: a Web application, a retrieval service, and one or more Generative Artificial Intelligence (GenAI) models. These components are all generic and substitutable. We provide demonstrable implementations of all these components in an end-to-end prototype. In practice, Web app developers and service providers would develop their own respective Apps or services that integrate easily with other components in the ecosystem.

In SocialGenPod, the user interacts with a Web app, which establishes the user's identity using Solid-OIDC. The app retrieves and stores user data such as configuration and chat history in their Solid Pod. The user may manually configure the app to select their preferred providers for retrieval and generative AI models. 

The retrieval and generative AI services may run externally. The generative AI services do not have direct access to the Web app user's Pod. This allows the user to easily switch between different service providers. We expect these services to be similar to Solid Pod services -- a marketplace of public or private, free or paid, fast or slower but high-accuracy services for the user to choose from (or self-host), depending on the quality of service, reputation, and trust.

On the other hand, based on the use case, the retrieval service will require access to the Pod it sources data from (which may or may not be the Web app user's Pod). This is subject to user-specified preferences such as which services can be used to access their data, which data, and for what purpose.

Restricted (e.g., private or paid subscription) model services can authenticate users using Solid-OIDC, which eliminates the need for self-developed account management mechanisms. For example, the model service can maintain an internal list of allowed users or read lists of trusted users from Solid Pods. It allows features like paid subscription and delegation of usage rights to friends.

While service providers could be deployed on any suitable infrastructure, it would be sensible for Solid Pod providers to offer such services alongside Pod hosting. As Solid Pod providers are already trusted by their users to store their data securely, this would eliminate the need to trust another third party. It would also boost performance, by potentially allowing data to be transferred between services running on the same host with low latency.

Although external services may cache user data for performance, this should be ephemeral and securely isolated. Since the service providers are expected to not store user data permanently, it is easier to establish trust and identify dishonest behaviours, thus encouraging trustworthy services.



\subsection{The Web App}

\begin{figure}[htb]
    \centering
    \includegraphics[width=\linewidth]{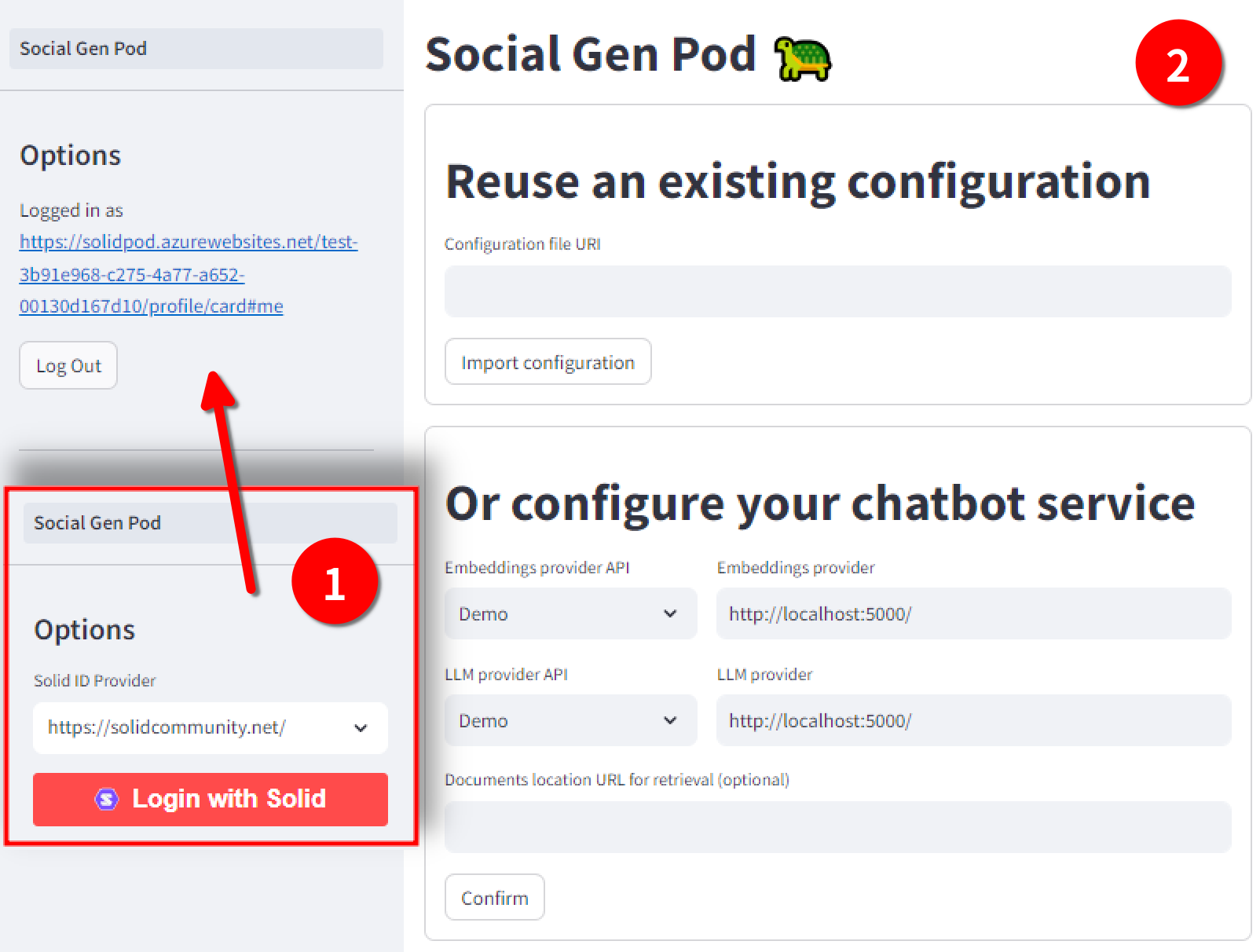}
    \caption{Login and configuration menu in SocialGenPod.}
    \Description[Screenshot with login button and configuration form prompting to reuse an existing configuration or configure your chatbot service]{2 step screenshot. Step 1 shows the app sidebar with the label ``Options'', a dropdown for Solid ID Provider with ``https://solidcommunity.net/'' selected, and a big button with the label ``Login with Solid''. Step 2 shows the user already logged in. On the right, there is a form prompting to reuse an existing configuration or configure your chatbot service. In the configuration options, the user can choose an API type for both a retrieval service provider and LLM provider -- currently ``Demo'' is selected for both. For each provider, the user can also input a service provider URL -- currently ``http://localhost:5000/''. Below, there is an optional text field to enter a location to source documents from.}
    \label{fig:setup-screenshot}
\end{figure}

\begin{figure}[h]
    \centering
    \includegraphics[width=\linewidth]{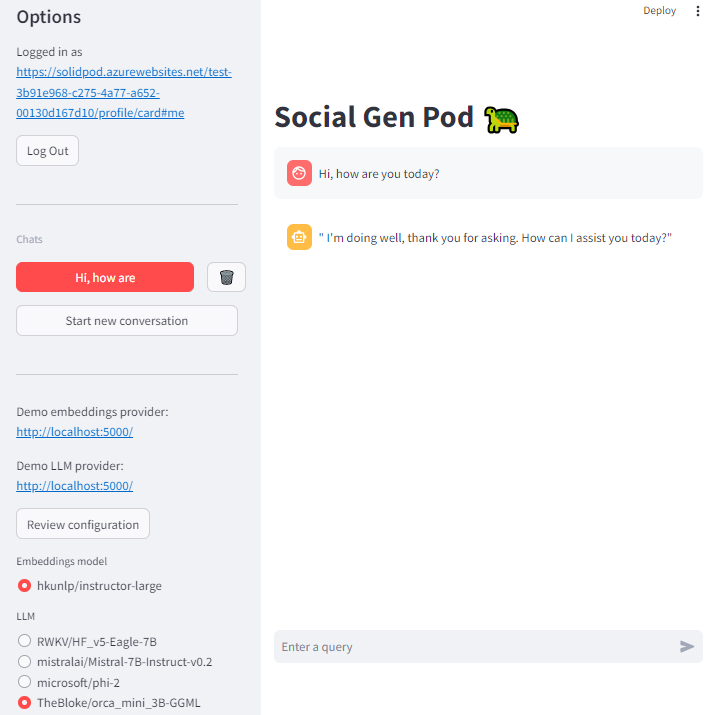}
    \caption{Full chat interface in SocialGenPod, showing chat threads and available models.}
    \Description[SocialGenPod sidebar and chat window]{A screenshot of SocialGenPod. On the left, a sidebar shows that the current user is logged in and there is a log out button. Below, there is a section called ``Chats'' with a selected button labelled ``Hi, how are". There is also a delete button and a ``Start new conversation'' button. Further below, it says Demo embeddings provider http://localhost:5000/ and Demo LLM provider http://localhost:5000/ and there is a button labelled ``Review configuration". Finally, there are radio buttons for embedding models and LLMs. On the right, a chat window shows a user question ``Hi, how are you today?'' and an AI response ``I'm doing well, thank you for asking. How can I assist you today?''}
    \label{fig:screenshot-sidebar}
    \vspace{-0.5cm}
\end{figure}



Our SocialGenPod implementation uses an embeddings model provider as the retrieval service and a Large Language Model provider as the generative AI model service.

Upon opening the SocialGenPod Web app, the user logs in using their Solid credentials (Step 1 in \autoref{fig:setup-screenshot}). Then, they can set up the service providers to use (or reuse a previous configuration) for retrieval and generation (Step 2 in \autoref{fig:setup-screenshot}). Once the app is configured, as shown in the left sidebar in \autoref{fig:screenshot-sidebar}, the user can choose to start a new conversation thread or continue an existing conversation. Chat threads and messages are stored in the user's Pod. Available models from selected providers are also shown in the sidebar, so the user can quickly switch between different models at any time, even mid-conversation.

The Web app handles data retrieval and updates from/to the user's Pod, and the language model service only receives conversation data from the Web app. When sending requests to external endpoints, the Web app sends Solid-OIDC authentication headers as well, so both the Solid Pod and the external services can confirm the identity of the requestor. In this way, the services can perform \emph{user} (but not \emph{account}) management, such as only allowing a predefined set of users to use the service.

Optionally, in the configuration step, the user may provide a Solid Pod location to source documents from for Retrieval Augmented Generation. If provided, the specified retrieval service gets relevant documents, which are sent together with user queries to the language model service. An alternative design choice could have been for the Web App to retrieve documents and send them to the retrieval service for similarity search, however, our chosen approach supports advanced social sharing, as discussed in \autoref{sec:social-sharing}.

\subsection{Social Sharing}
\label{sec:social-sharing}

Besides individual use cases, SocialGenPod supports social sharing of both service providers and personal documents.

The retrieval service can use Solid-OIDC for user identity verification. This way, Alice can selectively allow only Bob to read data from her Pod \emph{using a particular retrieval service}, independently of which front-end Web app Bob uses. Since the generative AI model provider is decoupled from the retrieval service provider, Bob could use any generative AI model in this workflow, self-hosted or remote.

The advantage of this micro-services-oriented plugin architecture is that it keeps services minimal and substitutable. Service providers do not need to maintain account management systems nor associated controlled data storage. As briefly discussed above, this reduces the complexity and vulnerability of these services, and could improve trust between service providers and end users.

\section{Limitations and Future Work}\label{sec:open}
With SocialGenPod, we demonstrated how one can build privacy-friendly generative AI Web applications using Solid. In future work, it would be useful add more personalisation features, such as fine-tuning on personal data and deploying personalised models in a decentralised way. The main challenges to address are delegating fine-tuning computation and inference on personalised models.

Another potential direction for future research could be addressing challenges related to document retrieval by relevance. Currently, SocialGenPod relies on a retrieval service provider copying personal documents for local computations, which requires trusting that the provider will handle data securely.

It would also be interesting to explore private inference techniques \cite{hao2022iron}, for improved privacy protection when using untrusted model providers. Other privacy and security risks related to personal AI assistants and methods to mitigate those risks should also be investigated in future work. For example, a malicious actor with access to a personal AI assistant could potentially attempt to retrieve the full contents of user's personal documents by prompting the assistant many times. 










\begin{acks}
This work was partially supported by Oxford Martin School's programme on ``Ethical Web and Data Architectures in the Age of AI''. We thank Tim Berners-Lee and Nigel Shadbolt for providing valuable feedback. 
\end{acks}

\bibliographystyle{ACM-Reference-Format}
\balance
\bibliography{sample-base}
\end{document}